# Network constraints in scale free dynamical systems

Erik D. Fagerholm[1,*], W.M.C. Foulkes[2], Yasir Gallero-Salas[3,4], Fritjof Helmchen[3,4], Karl J. Friston[5], Robert Leech[1,§], Rosalyn J. Moran[1,§]

[1] Department of Neuroimaging, King's College London
[2] Department of Physics, Imperial College London
[3] Brain Research Institute, University of Zürich
[4] Neuroscience Center Zürich
[5] Wellcome Trust Centre for Neuroimaging, University College London

[§] These authors contributed equally to this work

[*] Corresponding author: erik.fagerholm@kcl.ac.uk

**Abstract**

Scale free dynamics are observed in a variety of physical and biological systems. These include neural activity in which evidence for scale freeness has been reported using a range of imaging modalities. Here, we derive the ways in which connections within a network must transform – relative to system size – in order to maintain scale freeness and test these theoretical transformations via simulations. First, we explore the known invariance of planetary motion for orbits varying in size. Using parametric empirical Bayesian modelling and a generic dynamical systems model, we show that we recover Kepler's third law from orbital timeseries, using our proposed transformations; thereby providing construct validation. We then demonstrate that the dynamical critical exponent is inversely proportional to the time rescaling exponent, in the context of coarse graining operations. Using murine calcium imaging data, we then show that the dynamical critical exponent can be estimated in an empirical biological setting. Specifically, we compare dynamical critical exponents – associated with spontaneous and task states in two regions of imaged cortex – that are classified as task-relevant and task-irrelevant. We find, consistently across animals, that the task-irrelevant region exhibits higher dynamical critical exponents during spontaneous activity than during task performance. Conversely, the task-relevant region is associated with higher dynamical critical exponents in task vs. spontaneous states. These data support the idea that higher dynamical critical exponents, within relevant cortical structures, underwrite neuronal processing due to the implicit increase in cross-scale information transmission.

**Introduction**

The spatiotemporal patterns observed in certain systems, such as those exhibiting turbulent flow[1], appear to evolve with identical dynamics; regardless of the level of magnification at which they are observed – a property known as scale freeness[2]. The latter is of considerable interest in neuroscience due to increasing evidence that the brain exhibits scale freeness across several orders of magnitude; ranging from single-cell recordings[3], to meso-scale circuits[4] and entire brain regions[5]. Studies of scale freeness in neuroscience often address power law distributions in graph theoretic metrics[6] or the probability distributions of cascading events[7]. However, these metrics are often lacking in statistical methodology[8] and are therefore unable to rigorously characterize dynamics of different brain states.

A system operating with scale free dynamics exhibits a number of traits, such as divergence in correlation length[9] and fractal shape collapse[10]. These traits are in turn associated with functional benefits within neural systems, such as maximized information transmission and capacity[11]. Furthermore, scale freeness presents evolutionary and ontogenetic advantages, as the same neural architecture can be replicated across species with brains of different sizes (Fig. 1A), or across the lifespan of an individual animal (Fig. 1B)[12, 13, 14].

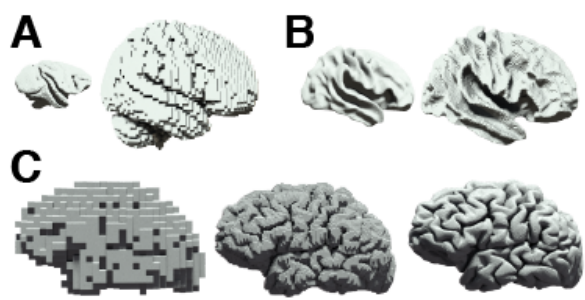

*Fig. 1: Scaled neural architectures* **A)** *Vervet monkey (left) and human (right).* **B)** *Inflated cortical surfaces from an infant (left) and adult (right) human.* **C)** *A human brain at three different levels of coarse graining.*

Scale freeness provides a theoretical framework in which findings at e.g. the level of individual synapses can be used to make cross-scale predictions. It is therefore of interest to establish links between the mathematical descriptions of scale free neural dynamics and the structures of networks in which they are constrained to operate. It is with this motivation in mind that we present a derivation of the ways in which the connections within a dynamical system must transform in order to maintain scale freeness. We proceed by developing a statistical framework to determine the scaling parameter values that best explain measured changes in dynamics across scales.

We begin with a known example from celestial mechanics that demonstrates the way in which Kepler's third law can be derived by applying scale freeness as a constraint in Newton's second law. We proceed by simulating orbits of increasing size and show that one can recover the Kepler scaling exponent from simulated data using a hierarchical statistical model, thus providing proof of principle. The hierarchical modelling uses variational Bayes to estimate the strength of effective connectivity between the orbiting bodies at each scale. This framework rests on the principles of Dynamic Causal Modelling (DCM) previously developed for the analysis of neuroimaging data, extended here with Parametric Empirical Bayes (PEB)[15] for the characterisation of scale free hierarchies.

We proceed by showing how scale freeness in a neural system can be quantified using coarse graining (Fig. 1C) together with an application of DCM and PEB. Specifically, we demonstrate the way in which the temporal rescaling exponent relates to the dynamical critical exponent as defined in Renormalization Group Theory[16]. Finally, we confirm predictive validity of this coarse graining approach by testing the hypothesis that there are discernable differences between the dynamical critical exponents associated with spontaneous activity and task performance across different neural regions. We use calcium imaging data collected in a murine model with high spatiotemporal resolution ($\sim 40\ \mu m, 50\ ms$). We find that the dynamical critical exponent is higher in spontaneous activity compared with task states in the forelimb, hindlimb, and motor cortices (task-irrelevant regions). The opposite relationship; i.e., a lower dynamical critical exponent in spontaneous activity compared with task states, is observed in the posterior parietal, visual, barrel, and trunk cortices (task-relevant regions).

**Methods**

**Scale free dynamical systems:** A dynamical system is said to be scale free if its Lagrangian $\mathcal{L}(q, \dot{q})$ transforms under change of scale as follows:

$$\mathcal{L}_s = k\mathcal{L}, \qquad [1]$$

where $\mathcal{L}_s$ is the Lagrangian describing the scaled system and $k$ is a constant[17].

We recover the equation of motion associated with the scaled system by following the principle of stationary action (Euler-Lagrange):

$$\frac{\partial \mathcal{L}_s}{\partial q} - \frac{d}{dt}\frac{\partial \mathcal{L}_s}{\partial \dot{q}} = 0, \qquad [2]$$

which, using [1], can be written as:

$$\frac{\partial [k\mathcal{L}]}{\partial q} - \frac{d}{dt}\frac{\partial [k\mathcal{L}]}{\partial \dot{q}} = 0, \qquad [3]$$

in which the constant $k$ cancels, leaving:

$$\frac{\partial \mathcal{L}}{\partial q} - \frac{d}{dt}\frac{\partial \mathcal{L}}{\partial \dot{q}} = 0. \qquad [4]$$

Therefore, the condition for scale freeness in a Lagrangian formulation via the introduction of a constant factor (equation [1]) is equivalent to requiring that the equation of motion describing the scaled system (equation [2]) be identical in form to the equation of motion describing the original unscaled system (equation [4]). It is this latter (non-Lagrangian) definition of scale freeness in the equations of motion describing dynamical systems that we use going forward.

**Scale freeness in celestial mechanics:** here we demonstrate how Kepler's third law follows from the property of scale freeness in Newton's second law. We use this as a motivating example of how scale freeness in the equations of motion can yield valuable information about the behaviour of a dynamical system.

The trajectory $r(t)$ of a planet orbiting a sun may be found via Newton's second law:

$$m\frac{d^2[r(t)]}{dt^2} = -\frac{GMm}{r^3(t)}r(t), \qquad [5]$$

where $m$ is the mass of the planet, $M$ is the mass of the Sun, and $G$ is the universal gravitational constant.

We now transform the planet's trajectory $r(t)$ to a scaled trajectory $r_s(t)$ as follows:

$$r(t) \rightarrow r_s(t) \triangleq br(b^\alpha t), \qquad [6]$$

where $b$ is an arbitrary scale factor and $\alpha$ is a constant to be determined.

In order to find the equation of motion satisfied by the scaled trajectory $r_s(t)$ we begin by replacing $t$ with $b^\alpha t$ in [5], such that:

$$m\frac{d^2[r(b^\alpha t)]}{d(b^\alpha t)^2} = -\frac{GMm}{r^3(b^\alpha t)}r(b^\alpha t), \qquad [7]$$

or equivalently:

$$m\frac{d^2[br(b^\alpha t)]}{dt^2} = -b^{2\alpha+3}\frac{GMm}{\left(br(b^\alpha t)\right)^3}br(b^\alpha t), \qquad [8]$$

which, using [6], can be written as:

$$m\frac{d^2[r_s(t)]}{dt^2} = -b^{2\alpha+3}\frac{GMm}{r_s^3(t)}r_s(t). \qquad [9]$$

The $b^{2\alpha+3}$ factor on the right-hand side prevents the scaled trajectory, $r_s(t)$, from satisfying Newton's second law in equation [5]. Instead, the scaled trajectory describes the motion of a planet orbiting a sun with a different mass: $M_s = b^{2\alpha+3}M$. However, if we choose $\alpha$ such that $M_s = M$, which occurs when:

$$2\alpha + 3 = 0 \implies \alpha = -{}^3\!/\!{}_2, \qquad [10]$$

then the equation of motion for the scaled trajectory $r_s(t)$ becomes identical to the equation of motion for the original trajectory $r(t)$. The value of $\alpha$ in [10] shows us that if $r(t)$ is a solution, then so is $r_s(t) \triangleq br(b^{-3/2}t)$ for any choice of scaling parameter $b$, or in other words that the square of the period of the orbit is proportional to the cube of its semi-major axis; i.e., Kepler's third law.

**Scale freeness in the DCM recovery model:** We now use the scale freeness in Newton's second law shown above as a ground-truth model in demonstrating that a generic DCM can be used to recover Kepler's third law from a simulation of orbital motion. First, we generate timeseries of two planets orbiting a sun using at different spatial scales. We then estimate the planetary gravitational interaction strengths via a recovery model in the form of a bilinear dynamical system approximation – that, crucially, can also be applied to arbitrary timeseries in which the true generative process is unknown.

We begin by using timeseries $r(t)$ that are solutions of equation [5] to generate orbital motion data. To recover the $\alpha$ scaling exponent from equation [6] we then assume that the planetary trajectories can be approximated by solutions of the bilinear form of the DCM recovery model:

$$\frac{d[r(t)]}{dt} = \left(A + \sum_j u_j(t) B^j\right) r(t) + C v(t) + \omega^{(x)}, \quad [11]$$

which is capable of modelling arbitrary dynamical systems beyond the context of planetary motion. Furthermore, the recovery model explicitly accommodates gravitational interaction strengths between the planets, but in principle can be used to estimate connectivity within and between nodes in a generic network. With reference to equation [11]: $r$ is a column vector representing (in the context of this orbital simulation) the distances of the two planets from the centre of gravity of the three-body (sun and two planets) system; $A$ is the intrinsic coupling matrix. Note that the reason this is called an *intrinsic* coupling matrix is that it mediates the influence of states on each other that are intrinsic to the system. In linear state space models this would be the system matrix that plays the role of a Jacobian. In neurobiology, the intrinsic coupling matrix is often referred to as an *average* matrix to avoid confusion between intrinsic (within a neuronal region) and extrinsic (between neuronal regions) connectivity; $u_j$ are priors on hidden causes; $\boldsymbol{v} = \boldsymbol{u} + \boldsymbol{\omega}^{(v)}$ where $\boldsymbol{\omega}^{(v)}$ is a noise term describing random, non-Markovian fluctuations on external perturbations[18] $\boldsymbol{u}$; $B^j$ are the bilinear coupling matrices; $C$ is the exogenous connectivity matrix; and $\boldsymbol{\omega}^{(x)}$ is an $n$-component column vector of noise terms describing random, non-Markovian fluctuations[19] on $\boldsymbol{r}$. If $\boldsymbol{r}$ has $n$ components and there are $m$ perturbing inputs $\boldsymbol{v}$, then $A$ and $B^j$ are $n \times n$ matrices, and $C$ is an $n \times m$ matrix. Although all numerical methods used here accommodate noise via $\boldsymbol{\omega}^{(x)}$, we omit this term for the sake of clarity in the presentation of the scaling theory below.

In order to retrieve the connectivity parameters from the recovery model in equation [11] we follow the same procedure as with Newton's second law and, as with the example described in equations [5] through [10], we seek ways in which to render the scaled equation of motion identical in form to the original unscaled equation of motion.

We begin by replacing $t$ with $b^\alpha t$ in [11], such that:

$$\frac{d[\boldsymbol{r}(b^\alpha t)]}{d(b^\alpha t)} = \left(A + \sum_j v_j(b^\alpha t) B^j\right) \boldsymbol{r}(b^\alpha t) + C\boldsymbol{v}(b^\alpha t), \qquad [12]$$

or equivalently:

$$\frac{d[b\boldsymbol{r}(b^\alpha t)]}{dt} = b^\alpha \left(A + \sum_j v_j(b^\alpha t) B^j\right) b\boldsymbol{r}(b^\alpha t) + b^{\alpha+1} C\boldsymbol{v}(b^\alpha t). \qquad [13]$$

Using [6], this can be written as:

$$\frac{d[\boldsymbol{r}_s(t)]}{dt} = b^\alpha \left(A + \sum_j v_j(b^\alpha t) B^j\right) \boldsymbol{r}_s(t) + b^{\alpha+1} C\boldsymbol{v}(b^\alpha t), \qquad [14]$$

which differs from [11] for all values of $\alpha$. Therefore, as opposed to Newton's second law, it is not possible to render the original equation [11] identical in form to the scaled equation [14] simply by specifying a value of $\alpha$. Instead, scale freeness (such that $\boldsymbol{r}_s(t) \triangleq b\boldsymbol{r}(b^\alpha t)$ becomes a possible solution) requires that the parameters of the DCM recovery model [11] also change relative to system size.

Specifically, we require that the frequency of external perturbations transform as follows:

$$\boldsymbol{v}(t) \rightarrow \boldsymbol{v}_s(t) = \boldsymbol{v}(b^\alpha t), \qquad [15]$$

which allows us to write [14] as:

$$\frac{d[\boldsymbol{r}_s(t)]}{dt} = b^\alpha \left( A + \sum_j v_{s,j}(t) B^j \right) \boldsymbol{r}_s(t) + b^{\alpha+1} C \boldsymbol{v}_s(t). \qquad [16]$$

Furthermore, the connectivity matrices must transform as follows:

$$A \rightarrow A_s = b^\alpha A, \qquad [17]$$

$$B^j \rightarrow B_s^j = b^\alpha B^j, \qquad [18]$$

$$C \rightarrow C_s = b^{\alpha+1} C. \qquad [19]$$

Using [17], [18] and [19] we can then write [16] as:

$$\frac{d[\boldsymbol{r}_s(t)]}{dt} = \left( A_s + \sum_j v_{s,j}(t) B_s^j \right) \boldsymbol{r}_s(t) + C_s \boldsymbol{v}_s(t), \qquad [20]$$

which we see is now identical in form to [11], thus achieving scale freeness.

Therefore, by using this DCM to recover connectivity parameters from ground truth data, we should be able to verify that the highest model evidence for the theoretically predicted relationships between connectivity and scale (equations [17], [18] and [19]) is obtained when the scaling exponent $\alpha$ lies close to the value $-3/2$, as known a priori from Kepler's third law (equation [10]).

**Orbital simulation:** We simulate three bodies orbiting a common centre of gravity using a modified version of a freely available n-body physics simulator as part of the Unity3D gaming engine[20] (version 2017.3.1f1). The mass of the star is $10^5$ times greater than that of the two planets. This is sufficiently massive such that the wobble of the star about the centre of gravity of the three-body system is zero to within-software precision. We begin with a simulation in which the semi-major axes of the orbits of the two planets differ by 10%. We run this simulation a total of ten times with the same initial conditions, except that in each new simulation we increase the sizes of both semi-major axes by 10%.

**First level modelling:** The time courses of the two planets are used to recover their intrinsic connectivity via Dynamic Expectation Maximization (DEM)[21] for each of the ten simulations. There are no perturbations arising from gravitational effects with other bodies – beyond the sun and two-planet solar system – and we therefore set the elements of the $B$ and $C$ matrices in equation [11] to zero. We initialised the model by setting planetary positions to unity at time $t = 0$. The DEM procedure seeks to estimate the 'true' signal (similar to a Kalman filter) but in doing so also provides an estimate of the intrinsic connectivity matrix, as well as an estimate of the hyperparameters; i.e., random fluctuations or noise in the states. The inversion uses a descent on variational free action over the time courses of simulated orbits and employs a Laplace approximation over states, parameters and hyperparameters to produce maximally entropic posterior probability distributions for any probabilistic model whose posteriors can be reasonably approximated with a Gaussian.

The inversion operates in generalised coordinates of motion, thereby accommodating non-smooth, non-Markovian noise processes. We use a prior variance of 1/64 and prior means of -2 for main diagonal and 0 for off-diagonal elements of the $A$ matrix. Furthermore, we use a hyperprior of 16 for the log precisions over observation noise for the orbit, amounting to a small noise assumption – due to the precision of the physics engine data.

**Second level modelling:** We then use a hierarchical PEB scheme to assess the degree to which changes in intrinsic connectivity matrix elements – recovered for different sized orbits – can be explained by the theoretically predicted transformation for scale free systems (equation [17]) for a range of scaling exponent $\alpha$ values. PEB was repeated for each plausible value of $\alpha$ (in steps of 0.01 between -3 and 0). This line search enabled us to track the free energy approximation to model evidence (as well as the posterior expectation of second-level parameters) as a function of the scaling exponent. Practically, we test equation [17] using the second level PEB model $Y = X\beta + \epsilon$, where $Y$ is a column vector comprising DCM estimates (means and variances) at different scales and $X$ is a column vector which contains the theoretically determined scale from equation [17], specified by the scaling parameter $b$. PEB returns the approximate model evidence (free energy), as well as the posterior over $\beta$, for each value of the scaling exponent $\alpha$. We are therefore in a position to test the hypothesis that the variation in connectivity with respect to scale in equation [17] is best explained using a scaling exponent value close to $\alpha = -3/2$ in accordance with Kepler's third law.

**Coarse graining and the DCM recovery model:** In the previous section we considered scaling operations in terms of changes to the size of a planetary system. However, in neuroimaging we face a different situation, in which data is collected at a single scale. In this setting, it no longer makes sense to think of scaling in terms of physical size. Instead, we now consider scaling in terms of changes to the resolution with which neuroimaging data is observed, via coarse graining.

We perform coarse graining throughout by repeating the following two steps as many times as an image will allow: 1) We combine $2 \times 2$ neighbouring 'regions' of an image into 'blocks', in which the time course of a given block is defined as the mean of the time courses of its constituent regions; 2) We then redefine regions such that each one now occupies the same spatial extent of the image as a block in step one. Similarly, we redefine blocks such that each one now consists of the newly defined larger $2 \times 2$ regions.

In the orbital mechanics example, we used the dependent variable $r$ to refer to the position of a given planet. In dealing with neuroimaging data, we now instead use $x$ and $X$ to refer to the measured signal intensities of a given region and block, respectively. Considering the way in which a given region's signal transforms with system size $b$ we must, in principle, alter equation [6] to include a new scaling exponent $\beta$ to account for non-linear changes to measured signal intensity:

$$x(t) \rightarrow x_s(t) \triangleq b^\beta x(b^\alpha t), \qquad [21]$$

as we have no reason to assume a linear change in measured signal intensity between averaged time courses from progressively larger portions of an image. However, in all subsequent analyses we deal with z-scored time courses with zero mean and unity variance, meaning that we can use the following simplified version of equation [21]:

$$x(t) \rightarrow x_s(t) \triangleq x(b^\alpha t). \qquad [22]$$

We can therefore say that a system is scale free if (on average) the following relationship between time courses of regions $x(t)$ and blocks $X(t)$ holds:

$$X(t) = x(b^\alpha t). \qquad [23]$$

It is this relationship that we test using the DCM recovery model (equation [11]). As with the orbital simulation we set the elements of the $B$ and $C$ matrices to zero, due to the fact that the neuroimaging data used throughout are unaffected by external perturbations.

**Intrinsic connectivity and the dynamical critical exponent:** Previously, we simulated orbital paths of different sizes and showed how the DCM recovery model can be used to estimate the scaling exponent $\alpha$ that encodes the way in which space and time scale relative to one another (equation [17]). However, as we are now analysing coarse grained neuroimaging data, we will instead focus on the relationship between block size and the ratio of characteristic relaxation times in blocks and their constituent regions. This relationship is encoded in the dynamical critical exponent $z$ as defined in Renormalization Group Theory. Below we show how the scaling exponent $\alpha$ relates to the dynamical critical exponent $z$ and how the latter can be estimated using the DCM recovery model.

We begin by defining a given region's characteristic decay time $t_r$ as the time taken for the following quantity:

$$\sigma_r^{\,2}(t) = \frac{1}{N}\sum_{i=1}^{N} x_i^{\,2}(t) = \langle x_i^{\,2}(t)\rangle_r \qquad [24]$$

to decay to $1/e$ of its initial value:

$$\sigma_r^{\,2}(t_r) = \sigma_r^{\,2}(0)/e, \qquad [25]$$

where the angle brackets in [24] with subscript $r$ denote averages over regions at a specific time.

Similarly, we define a given block's characteristic decay time $t_b$ as the time taken for the following quantity:

$$\sigma_b^{\,2}(t) = \frac{b^2}{N}\sum_{I=1}^{\frac{N}{b^2}} X_I^{\,2}(t) = \langle X_I^{\,2}(t)\rangle_b \qquad [26]$$

to decay to $1/e$ of its initial value:

$$\sigma_b^{\,2}(t_b) = \sigma_b^{\,2}(0)/e, \qquad [27]$$

where the angle brackets in [26] with subscript $b$ denote averages over blocks at a specific time.

If the system is scale free, then using [23], [24] and [26] we obtain:

$$\sigma_b^{\,2}(t) = \sigma_r^{\,2}(b^\alpha t). \qquad [28]$$

If we choose to set $t = b^{-\alpha} t_r$, then using [25], [27] and [28] we see that:

$$\sigma_b^{\,2}(b^{-\alpha} t_r) = \sigma_r^{\,2}(t_r) = \frac{\sigma_r^{\,2}(0)}{e} = \frac{\sigma_b^{\,2}(0)}{e}, \qquad [29]$$

where $\frac{\sigma_b^{\,2}(0)}{e}$ is the definition of the block time scale $t_b$ from [27] and hence:

$$t_b = b^{-\alpha} t_r. \qquad [30]$$

The dynamical critical exponent $z$ is defined as follows:

$$t_b = b^z t_r, \qquad [31]$$

which, together with [30], shows us that:

$$\alpha = -z. \qquad [32]$$

Therefore, in the context of coarse graining, by estimating $\alpha$ in [17] we are in fact estimating the negative dynamical critical exponent, such that the transformation of intrinsic connectivity in [17] can be re-formulated as:

$$A \rightarrow A_s = b^{-z} A. \qquad [33]$$

It is this relationship that we test in the analysis of neuroimaging data, in order to recover the dynamical critical exponent that best explains the measured changes in connectivity under progressive coarse graining operations.

**Animals and surgical procedures:** All animal experiments were carried out according to the guidelines of the Veterinary Office of Switzerland following approval by the Cantonal Veterinary Office in Zürich.

We use 3 triple transgenic Rasgrf2-2A-dCre; CamK2a-tTA; TITL-GCaMP6f adult male mice (3-5 months old). This line is characterised by inducible, specific and high expression of the calcium indicator GCaMP6f in pyramidal layer 2/3 neurons of the neocortex[22]. To induce the expression of the indicator, destabilized Cre must be stabilized by trimethoprim (TMP). Individual mice are intraperitoneally injected with 150 µg TMP/g of body weight reconstituted in Dimethyl sulfoxide (DMSO, Sigma 34869) at a saturation level of 100 mg/ml.

In order to expose the skull above the left brain hemisphere for wide-field calcium imaging, we use the minimally invasive intact skull preparation described previously[23]. Briefly, mice are anaesthetized (2% isoflurane in pure O2) and their temperature controlled (37°C). After removing the skin and connective tissue above the dorsal skull, we clean and dry the skull. We then apply a layer of UV-cure iBond over the skull, followed by a second layer of transparent dental cement (Tetric EvoFlow T1). Dental cement "worms" (Charisma) are applied around the preparation and a metal head post for head fixation is glued to the preparation. The resulting imaging window ranges from ~3 mm anterior to bregma to ~1 mm posterior to lambda and ~5 mm laterally to midline.

**Wide-field calcium imaging:** Calcium dynamics over the whole dorsal cortex of the left hemisphere are recorded using a wide-field imaging approach. Excitation light emanates from a blue LED (Thorlabs; M470L3) and is filtered (excitation filter, 480/40 nm BrightLine HC), diffused, collimated and directed to the left hemisphere by a dichroic mirror (510 nm; AHF; Beamsplitter T510LPXRXT). The imaging system consists of two objectives (Navitar, top objective: D-5095, 50 mm f0.95; bottom objective inverted: D-2595, 25 mm f0.95). Excitation light is focussed approximately 100 $\mu$m below the blood vessels. Green emission photons are collected through both objectives and dichroic, filtered (emission filter, 514/30 nm BrightLine HC) and recorded with a sensitive CMOS camera (Hamamatsu Orca Flash 4.0) mounted on top of the system. No photobleaching is observed under these imaging conditions. Images of 512x512 pixels are collected at 20 frames per second.

**Sensory mapping and alignment:** In order to align brain areas to the Allen Mouse Common Coordinate Framework[24], we perform sensory mapping under light anaesthesia (1% isoflurane) in each mouse. We present five different stimuli contralateral to the imaging side: a vibrating bar coupled to a loudspeaker is used to stimulate either 1) whiskers; 2) forelimb paw; or 3) hindlimb paw (somatosensory stimuli; 20Hz for 2s); 4) a 2s-long white noise sound is played (auditory stimulus); and 5) a blue LED positioned in front of the right eye provides a visual stimulus (100ms duration; approximately zero elevation and azimuth). The stimuli activate a corresponding set of cortical areas. These areas, together with anatomical landmarks (Bregma; Lambda; midline; as well as the anterior, posterior, and lateral ends of the dorsal cortex) are used as anchoring points to align each individual brain to the Mouse Common Coordinate Framework. Pixels outside the borders of the Mouse Common Coordinate Framework are discarded.

**Behavioural task:** Water-deprived head-fixed mice are trained in a go/no-go auditory discrimination task with a delay. Each trial (10s duration) commences with a trial cue (visual cue delivered by an orange LED, 1flash, 500ms duration) after which mice had to discriminate between two auditory tones (4 versus 8 kHz) presented for 2s. After a delay period (2-3s) a reward cue (3 flashes, 150ms duration with 100ms interval) signals the start of the response window (2s). Pure auditory sounds are generated by a Tucker-Davis System 3 processor (RZ6) and are presented using a magnetostatic loudspeaker (MF-1, Tucker-Davis) placed ~5cm from the right ear (contralateral to the imaged hemisphere). Each trial is separated by an inter-trial interval of ~5s.

Mice are trained using the 8 kHz tone as the 'go' stimulus. In order to obtain a water reward, mice have to lick a water spout in the response window during go trials ('hit'). Licks in response to the 'no-go' tone are mildly punished with white noise and a time out (~2s, 'false alarms', FA). Licks outside the response window ('earlies') are equally punished. The absence of licks in 'no-go' ('correct-rejections', CR) and 'go' ('misses') trials are neither rewarded nor punished. Performance is quantified as d-prime[25]: d' = Z(Hit/(Hit+Miss)) – Z(FA/(FA+CR)) where Z denotes the inverse of the cumulative distribution function. Animals are imaged upon reaching expert level performance ($d' > 1.5$), specifically $d'$=1.90, 2.23, and 2.36 for mouse 1, 2 and 3, respectively.

**Spontaneous activity:** We record meso-scale spontaneous activity in the same three mice that are imaged solving the task (using the same wide-field set-up) with equal trial and inter-trial interval lengths. Calcium dynamics are recorded in the absence of any external stimuli with the exception of a continuous blue light used for wide-field imaging (also present during task). This light is directed into the intact skull preparation from the optical path (placed above the heads of the mice) with an illumination intensity of < 0.1 mW/mm$^2$. The light in the recording environment is dim as the light is collimated and the objective is close to the preparation.

**Movement:** Although the animals are head-fixed they are able to freely whisk and move their limbs and backs. Given the dim recording conditions, we use infrared light to monitor the animals' movements (940nm infra-red LED) in both states (task and spontaneous activity). We extract movement vectors of the forelimb and back region from the recordings. Movement is calculated as 1 minus frame-to-frame correlation of these two regions. We perform multiple linear regression of all recordings with respect to the animals' movements, as well as to the external stimuli (sound and light cues) in the task recordings. It is due to this regression of movement and stimuli that we set the elements of the $B$ and $C$ matrices in the DCM recovery model to zero.

**Data pre-processing:** Matlab software (Mathworks) was used to pre-process the data. 512x512 pixel images are collected with the wide-field system and then downsampled to 256x256. Pixel size after downsampling was ~40 $\mu$m. To normalize for uneven illumination or GCaMP6f expression, we calculated the percentage change of fluorescence (ΔF/F) relative to the start of each trial.

**Regions of interest:** We begin by defining two non-overlapping regions of interest (ROIs) within the Allen Mouse Common Coordinate Framework that each span $64 \times 64$ pixels, as this is the largest power of $2$ that can be accommodated within the imaged area. The first ROI covers principally (as designated by the Allen Institute) primary somatosensory areas upper and lower limb. It also includes parts of the primary and secondary motor areas; primary somatosensory area unassigned; primary somatosensory area trunk; primary somatosensory area barrel field; and the retrosplenial area. The second ROI covers principally the posterior parietal association areas; the anteromedial visual area; and the posteromedial visual area. It also includes parts of the primary visual area; primary somatosensory area barrel field; and primary somatosensory area trunk. Data outside the ROIs are disregarded.

**Coarse graining**: Note that we use the term 'ROI' to refer to the 2 large areas defined above, whereas we use the term 'region' to refer to the constituents of 'blocks' in the language of Renormalization Group Theory. For each of the two ROIs, we then: a) z-score each region's time course in the $64 \times 64$ ROI; i.e., we subtract the mean and divide by the standard deviation on a region-wise level; b) subdivide the $64 \times 64$ ROI into a grid consisting of $32 \times 32$ blocks; c) run first level DCM on each of the $32 \times 32$ blocks, in which all connectivity matrices entered into first level DCMs are of size $2 \times 2$; d) perform Bayesian model averaging on the $32 \times 32$ first level DCMs such that we obtain a single representative intrinsic connectivity matrix ($A$ in equation [17]) associated with the first scale; e) coarse grain the $64 \times 64$ regions by a factor of 2 such that we obtain $32 \times 32$ regions, each of which corresponds to the mean of a $2 \times 2$ block within the original $64 \times 64$ ROI.

We then repeat steps a) through c) above for $32 \times 32$, $16 \times 16$, $8 \times 8$, $4 \times 4$ and $2 \times 2$ regions, each time recovering the intrinsic connectivity matrix associated with each level of coarse graining, where a quarter of the blocks are randomly sampled in step c) above for the first three scales in the interest of computational expediency. Note that in this characterisation of coupled dynamics we are averaging over different combinations of regions at any given scale. In other words, we are only interested in the coupling strengths that are conserved over regions (and not structured coupling between different regions at any given scale).

With reference to step d) we use a prior variance of $1$, and prior means of -1 for the main diagonal and $0$ for the off-diagonal coupling parameters of the $A$ matrix. In other words, we are, a priori, assuming each region can be positively or negatively influenced by any other region, while maintaining dynamical stability via self-inhibition. We use a hyperprior of 5 for the log precisions over observation noise, amounting to a larger noise assumption than for the orbital simulation.

We then enter the intrinsic connectivity matrices recovered at each level of coarse graining into the second level of the hierarchical modelling (PEB). We compare each scale to the original full-resolution $64 \times 64$ region data and test the extent to which the theoretical transformation in equation [33] holds. We run PEB for a range of dynamical critical exponents in steps of 0.005 between 0.1 and 0.4 in order to find the exponent associated with the highest model evidence (i.e., free energy).

**Results**

**Orbital simulation:** here we simulate the motion of two planets orbiting a star with progressively scaled orbital paths (Fig. 2A). The semi-major axes of the orbits are progressively increased, and Bayesian model inversion was used to recover the intrinsic connectivity matrices associated with each scale (Fig. 2B).

We then used PEB to assess how well the variation in intrinsic connectivity across scales is accounted for by the theoretical transformation in equation [17] for a range of power law exponents $\alpha$. The peak log model evidence (a.k.a., marginal likelihood approximated by free energy) for the entire two-body system (Fig. 2C, last column on right) was found at $\alpha = -1.47$, i.e., close to the value of $\alpha = -{3}/{2}$ as expected from Kepler's third law.

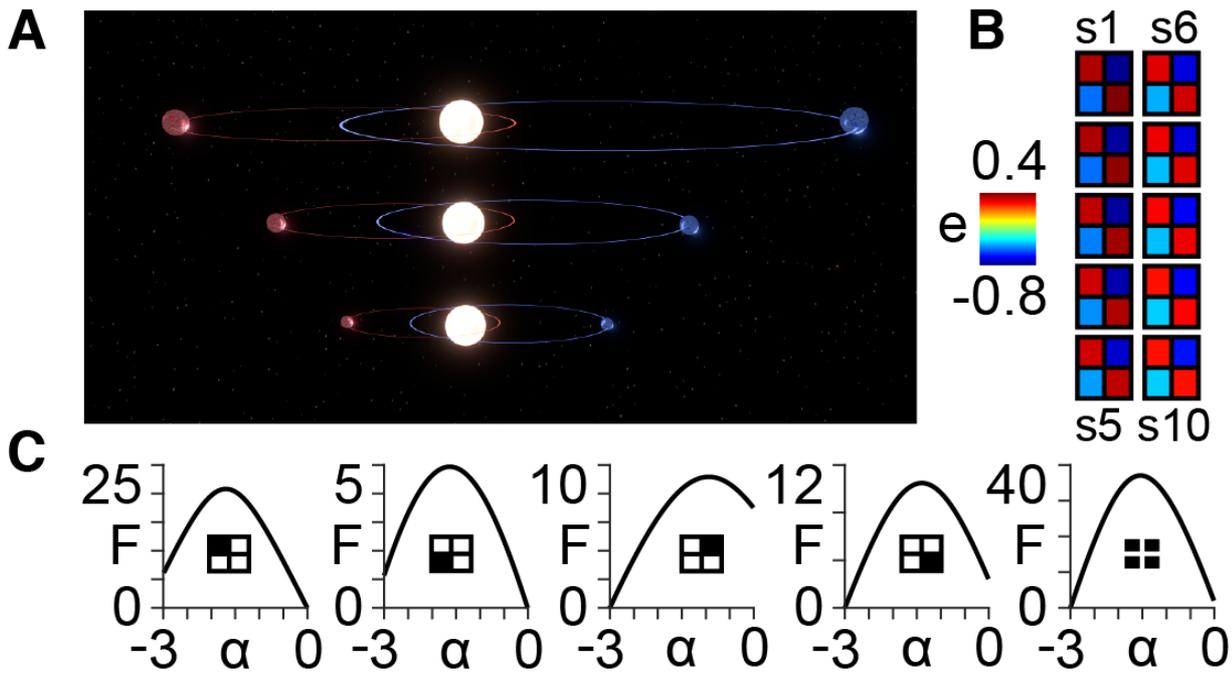

**Fig. 2: Orbital simulation (A)** The two-planet system orbiting a star at three different scales **(B)** A posteriori estimates (e) of coupling strengths following first-level modelling of the two-planet system for each of the ten orbital scales (s1 to s10). **(C)** Approximate lower bound log model evidence given by the free energy following second-level modelling of the ten scales shown in (B) as a function of the power law exponent $\alpha$. The first four panels (from left to right) pertain to the individual intrinsic coupling matrix elements shown in the insets. The fifth column shows the results summed across the four individual matrix elements.

**Neuroimaging data:** here we use a coarse graining approach to test the degree of scale freeness in calcium imaging data collected in mice (Fig. 3A & B) that are either awake and at rest (spontaneous activity) or performing a task. Results are presented for $n = 3$ mice, with analyses performed separately within two ROIs (Fig. 3C). ROI 1 (the top white square in Fig 3B) covers principally forelimb, hindlimb, and motor cortices; i.e., areas not directly involved in the task. ROI 2 (the bottom white square in Fig 3B) covers principally posterior parietal and visual; i.e., areas directly involved in the task.

We note three main results with reference to Figure 3C, which were remarkably consistent over the three mice analysed. Firstly, all values of the dynamical critical exponent $z$ are positive, which indicates (via equation [31]) that signal fluctuations decay more slowly in larger cortical structures.

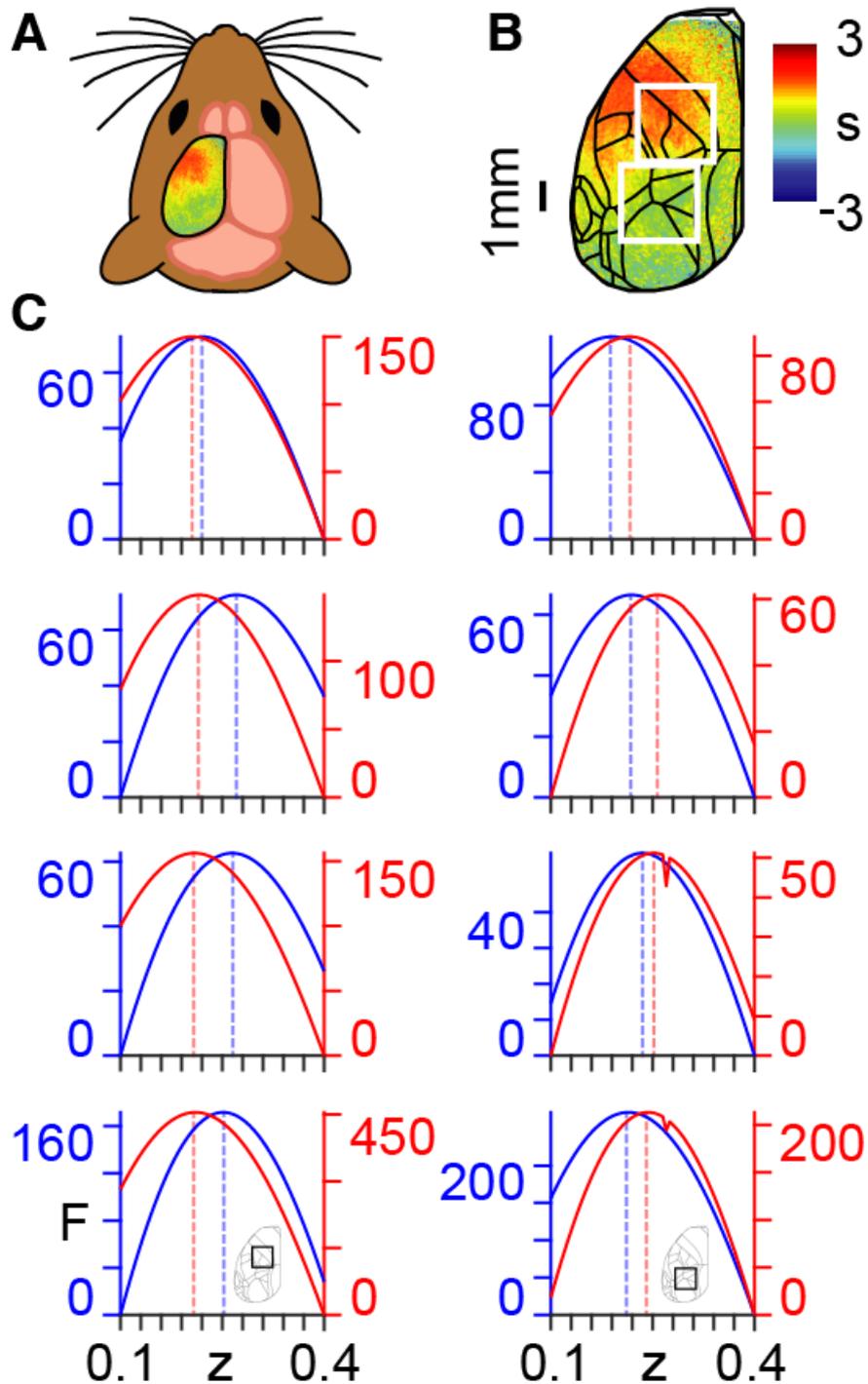

*Fig. 3: Coarse graining of calcium imaging data:* **A)** Wide-field calcium imaging over the left hemisphere of (three) head-fixed mice, expressing GCaMP6f in layer 2/3 excitatory neurons. **B)** Example z-scored (DF/F) activity averaged over a 10s trial length, shown as standard deviation (s) of the signal from the mean. Cortical areas are aligned to the Allen Mouse Common Coordinate Framework. The top and bottom white squares correspond to ROIs 1 and 2, respectively. **C)** Approximate lower bound log model evidence given by the free energy (F) as a function of the dynamical critical exponent (z) following PEB modelling across coarse-grained scales for spontaneous (blue) and task (red) states (summed across 'hit' and 'CR' trials), with maximum values indicated by the dashed vertical lines. Results in the left and right columns correspond to ROIs 1 and 2 in B, respectively, indicated also by the insets in the bottom row. Free energy values are presented individually for the three mice (rows 1-3 from top to bottom) and summed across the three mice (row 4, bottom).

Secondly, comparing across brain states within an ROI, we find that the dynamical critical exponent $z$ – that maximises marginal likelihood (i.e., model evidence or free energy) – is higher across all three animals in the spontaneous state, as compared with the task state in ROI 1 (task-irrelevant). Conversely, the dynamical critical exponent, $z$, corresponding to the maximum free energy is higher across all three animals in the task state as compared with the spontaneous state in ROI 2 (task-relevant). Thirdly, comparing across ROIs within a brain state, the dynamical critical exponent, $z$, corresponding to the maximum free energy in the spontaneous state is lower in ROI 2 (task-relevant) relative to ROI 1 (task-irrelevant) across all three animals. On the other hand, the dynamical critical exponent, $z$, corresponding to the maximum free energy in the task state is higher in ROI 2 (task-relevant) relative to ROI 1 (task-irrelevant) across all three animals.

**Discussion**

In order to quantify the degree of scale freeness one can use so-called box counting approaches – the coastline of the United Kingdom being a famous application thereof[26]. Box counting entails repeating the procedure of tracing the perimeter of the object in question with a small box shape, doubling the size of the box and then re-tracing. One then obtains a measure of the degree of scale freeness of the object by plotting the length of the perimeter as a function of box size. This is essentially the same process that we have followed by applying hierarchical DCM either across different spatial scales (as with the orbital simulation), or across different levels of coarse graining (as with the calcium imaging data).

We begin with the assumption that we are observing a phenomenon that possesses some degree of scale freeness. We then show how one can obtain a parameterisation of the relationship between scale and DCM coupling. These theoretically predicted relationships can then be used as a forward generative model in estimating the exponent of the power law relationship between a) space and time ($\alpha$ in equation [6]) for systems that are progressively scaled in size, and b) block size and the dynamical critical exponent ($z$ in equation [31]) for images that are progressively coarse grained.

We first evaluate posteriors over $\alpha$ in equation [6] in a simulation of planetary orbits, which serves as a ground-truth model in which the value of the scaling exponent is known from Kepler's third law. We apply nonlinear first level DCM at each orbital scale to obtain a series of intrinsic connectivities via equation [17]. We proceed by using linear second level PEB models to test the extent to which scale-related changes in connectivities can be explained across a range of power law exponents $\alpha$. We find that the highest model evidence is obtained close to $\alpha = -3/2$, as expected from Kepler's third law; thereby establishing face validity of this kind of estimation scheme.

As we have empirical evidence for scale freeness in neural systems[27, 28, 29], we use a similar technique in neuroimaging data. Specifically, we demonstrate how the parameterisation of the relationship between scale and DCM coupling can be altered to allow for the analysis of progressively coarse-grained images. We show that the degree of coarse graining relates to the physical size of a system via the dynamical critical exponent (see equation [32]). We therefore obtain a theoretical transformation of intrinsic connectivity under coarse graining operations; that allows for a measurement of the dynamical critical exponent (see equation [33]).

We considered two ROIs. ROI 1 covers mainly forelimb and hind limb somatosensory areas (tactile sensation), which are unlikely to be necessary for task performance. ROI 2 covers mainly the posterior parietal cortex – an area known to be involved in auditory decision making[30] and therefore likely to be necessary for task performance. We then note the following two points: a) the dynamical critical exponent with the maximum model evidence is higher across all three animals in the spontaneous state, as compared with the task state in ROI 1 and vice versa for ROI 2; b) the dynamical critical exponent with the maximum model evidence in the spontaneous state is higher in ROI 1 relative to the spontaneous state in ROI 2 across all three animals, and vice versa for the task state. Higher dynamical critical exponents reflect a greater degree of temporal renormalization across spatial scales (see equation [31]). The higher task-state exponents in ROI 2 (task-relevant) could be functionally beneficial due to an increase in cross-scale communication, via the temporal renormalization of signals within cortical structures engaged in the ongoing task. The lower task-state exponents in ROI 1 (task-irrelevant) could also be advantageous in terms of neuronal processing, as the associated reduction in temporal renormalization acts as a suppression mechanism of cross-scale message passing between cortical structures not involved in the task.

There are a range of scenarios in biology in which one may wish to characterize scale free systems, for example in the phylogenetic or ontogenetic scaling of neural structures. In neuroimaging one would commonly account for differences in scale by first projecting data into a common space before beginning the analysis of neural dynamics. For example – when comparing across development – neonatal, child and adult brains are first aligned onto a common template. Similarly, when comparing different species such as rodents, primates and humans, homologues are first identified between brain regions. The techniques we propose here present a novel quantification of the way in which network connectivities change in proportion to brain size, as well a method of constructing generative models for the characterisation of scale freeness within a formal statistical framework.

**Acknowledgements**

E.D.F. and R.L. were funded by the Medical Research Council (Ref: MR/R005370/1); K.J.F. was funded by a Wellcome Principal Research Fellowship (Ref: 088130/Z/09/Z); R.J.M. was funded by the Wellcome/EPSRC Centre for Medical Engineering (Ref: WT 203148/Z/16/Z).

**Author contributions**

Y.G.S. and F.H. collected the murine calcium imaging data; All authors designed and performed research, analysed data and wrote the paper.

**Competing interests**

The authors declare no competing interests.